\documentclass{article}
\usepackage[english]{babel}

\usepackage[letterpaper,top=2cm,bottom=2cm,left=3cm,right=3cm,marginparwidth=1.75cm]{geometry}
\usepackage{authblk}
\usepackage[numbers,comma,sort&compress]{natbib}
\usepackage{hypernat}
\usepackage{amsmath}
\usepackage{graphicx}
\usepackage[colorlinks=true, allcolors=blue]{hyperref}
\usepackage{placeins} 
\usepackage{tabularx}
\usepackage{tabu}
\usepackage{booktabs}
\usepackage{array}
\usepackage{makecell}
\usepackage{siunitx}
\usepackage{ulem}

\usepackage{multicol}

\newcommand{\LSMO}{La$_{0.7}$Sr$_{0.3}$MnO$_3$}
\newcommand{\LCMO}{La$_{0.7}$Ca$_{0.3}$MnO$_3$}
\newcommand{\YBCO}{YBa$_2$Cu$_3$O$_7$}

\newcommand{\etal}{\textit{et al}. }
\newcommand{\ie}{\textit{i}.\textit{e}. }

\title{Fabrication of planar halfmetallic ferromagnetic Josephson junctions with long range coupling}
\author{Junxiang Yao }
\author{Jan Aarts\footnote{Author to whom correspondence should be addressed: aarts@physics.leidenuniv.nl}}
\affil{Huygens-Kamerlingh Onnes Laboratory, Leiden Institute of Physics, \\ Leiden University, P.O. Box 9504, 2300 RA Leiden, Netherlands}

\date{}

\begin{document}

\maketitle

\begin{abstract}
Superconducting junctions with a ferromagnet as the weak link, where triplet correlations can transport supercurrents over a substantial distance, have been of long-standing interest. In this work, we study the triplet transport in planar \LSMO (LSMO) nanowire Josephson junctions with NbTi superconducting contacts. By meticulous ion etching with an artificial Pt hard mask, the NbTi/LSMO bilayer is structured to form an LSMO bridge without damaging its top layer. We observe superconducting (critical) currents of the order of 10$^{9}$~A/m$^2$ in a junction with a length of 1.3~$\mu$m, and distinguishing superconducting quantum interference (SQI) patterns when sweeping a magnetic field perpendicular ($B_\perp$) to the plane of the wire or parallel ($B_\parallel$) to the plane and along the wire. The observed Gaussian-shaped SQI pattern is attributed to the diffusive transport of triplet pairs in the LSMO. Our work demonstrates that combinations of oxide magnets with conventional ($s$-wave) alloy superconductors can be a promising new route to realizing superconducting spintronics.
\end{abstract}

\begin{multicols}{2}


In Josephson junctions of two superconductors (S) with a halfmetallic ferromagnetic (HMF) weak link, it is possible to convert conventional (spinless) Cooper pairs to equal-spin triplets that can in principle carry supercurrents over micrometer distances. This long range proximity (LRP) effect stems from the fully spin-polarized nature of the HFM, in which the triplets are not susceptible to pairbreaking exchange fields. The singlet-triplet conversion requires some form of spin mixing at the S/HMF interface, which can be in the form of magnetic inhomogeneities, or engineered by inserting additional ferromagnetic (F) layers. Long range effects have been reported in various experiments on devices with planar geometries \cite{keizer2006spin,anwar2010long,singh2016high,sanchez2022extremely}. Triplets were generated in CrO$_2$-based hybrids via magnetic inhomogeneities established either by strain or disorder of grain boundaries \cite{2011anwar}, while high-density critical supercurrents were observed using stacked S/F/CrO$_2$ geometries \cite{singh2016high}. Similarly, the LRP effect was studied in combinations of the high-T$_c$ superconductor \YBCO (YBCO) and the magnetic manganites \LCMO (LCMO) and \LSMO (LSMO) \cite{sanchez2022extremely,visani2012equal}. Visani \etal reported equal-spin Andreev reflections at the interface between YBCO and LCMO in a vertical architecture, as well as the existence of triplets in the LCMO layer of 30~nm \cite{visani2012equal}. Interestingly, uncompensated magnetic moments were discovered at the Cu-O-Mn chains across the interface between YBCO and LCMO that could be a source of magnetic inhomogeneity, although the relation to triplet generation was not elucidated \cite{chakhalian2006magnetism}. Very recently,  extremely long coherent transport of triplets was found in lateral YBCO/LSMO/YBCO junctions \cite{sanchez2022extremely}. Sanchez-Manzano \etal observed LRP effects in quite wide (20 or 25~$\mu$m) but also quite long (1~$\mu$m) LSMO wires, where multiple magnetic domain walls were present and possibly played a role in triplet generation and transport. Despite the LRP effect being unequivocally observed in the high-T$_c$ S/HMF systems, the underlying physics, in particular with respect to the triplet generation, still remains to be explored. \\

To further study the triplets generation and transport in HMF, in this work we fabricated lateral junctions of LSMO with the conventional superconductor NbTi, by carefully ion etching $in~situ$ grown NbTi/LSMO bilayer using artificial Pt hard masks. By starting from a bilayer, we avoided contamination of the LSMO/NbTi interface as it would occur when transfering the LSMO film $\it{ex}$-$\it{situ}$ for further processing. NbTi does not grow epitaxially on LSMO, differing from the high-$T_c$ S/HMF systems, but we know from earlier work that triplets are generated at the NbTi/LSMO interface \cite{jungxiang2023triplet}. The removal of the NbTi on LSMO was examined by energy-dispersive X-ray spectroscopy (EDS) characterizations and electrical resistance measurements, confirming the LSMO-based junctions. We observed the LRP effect in a junction with 1.3~$\mu$m length and 5~$\mu$m width. The critical current of this junction reached 11$\times$10$^8$~A/m$^2$ at 1.5~K. By sweeping the magnetic fields $B_\perp$ (perpendicular to the plane of the wire) and $B_\parallel$ (field in-plane and along the wire) at 6.3~K, just below the temperature where zero resistance is reached, distinguishing superconducting quantum interference (SQI) patterns were obtained. We attribute those Gaussian-shaped SQI patterns to the diffusive transport of triplets in LSMO. Our work demonstrates that strong triplet supercurrent can be found in conventional ($s$-wave) alloy S/HMF junctions. This may shed light on understanding the generation and transport of triplets in HMFs, and also offers a platform for spintronics-type applications.
%
%
\begin{figure*}[ht]
\centering
\includegraphics[width=.75\textwidth]{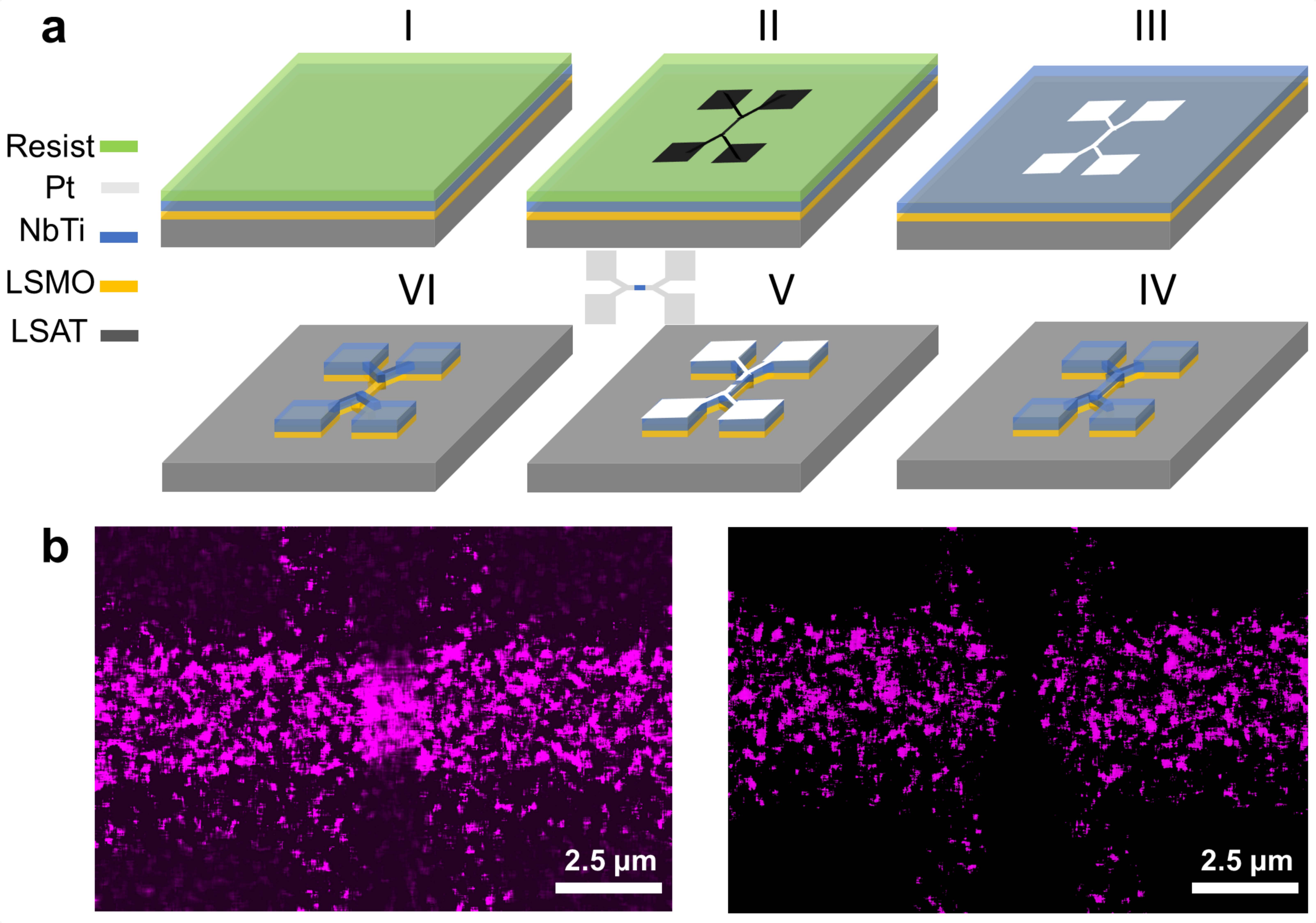}
\caption{\label{lrp-fig1} (a) Illustrative scheme of the nanofabrication procedure of NbTi/LSMO junctions. I: A positive resist is coated on NbTi/LSMO bilayer. II: The desired pattern is written in the positive resist by electron beam lithography (EBL). III: Pt is deposited and becomes a mask after removing the positive resist. IV: The NbTi/LSMO bilayer is patterned by ion beam etching (IBE) with a Pt mask. Note that, ideally, the Pt mask should be removed completely in this step. V: Repeating steps I to III, a second Pt mask is made again to further structure the patterned NbTi/LSMO bilayer by leaving the central region uncovered (see the small sketch with top view; the blue part is the uncovered region). VI: The NbTi/LSMO lateral junction is fabricated by employing IBE, and the central region is an LSMO (yellow) nanowire without NbTi on top. (b) EDS analyses of the Nb in a junction with a length of 1.1~$\mu$m before etching (left) and after the optimal etching (right), showing the removal of Nb in the gap area. The signal derives from the Nb K$_{\alpha}$ line at 16.6 keV, taken with a beam energy of 30~keV.}
\end{figure*}

\begin{table*}[t]
\begin{minipage}{\textwidth}
    \centering
    \caption{ Overview of experimental parameters in steps IV and VI in the nanofabrication process. The thickness of the Pt mask is adapted to the different etching requirements.}
    \label{lrp-etching}
        \begin{tabular}[t]{lS[table-format=2]*{5}{S[table-format=1]}}
            \toprule
            & \multicolumn{3}{c}{Step IV} & \multicolumn{2}{c}{Step VI}\\
            \cmidrule(r){2-4} \cmidrule(l){5-6}
            & {\thead{thickness(nm)}}
              & {\thead{etch rate(nm/min)}}
                & {\thead{etch time(min)}}
                  & {\thead{thickness (nm)}}
                    & {\thead{etch time(min)}}
                      \\
           \midrule
            Pt & 426  & 25.9 & 16.5 & 255.5 & 8 $\sim$ 9 \\
            NbTi  & 60  & 7.5 & \sim 0 & 60 & 8 $\sim$ 9 \\
           LSMO & 40 & 6.67 & \sim 0 & 40 & \sim 0 \\
            \bottomrule
        \end{tabular}
    \end{minipage}
\end{table*}
%



 We fabricated the lateral NbTi/LSMO junctions by carefully and accurately etching the NbTi/LSMO bilayer with an artificial Pt hard mask. By starting from a bilayer, the interface between NbTi and LSMO is not damaged. The details of the nanofabrication process are schematically depicted step by step in Fig.~\ref{lrp-fig1}a. \\
 \noindent
 The starting point for the fabrication is growing a NbTi/LSMO bilayer. The LSMO thin film (40~nm) was deposited on a (001)-oriented LSAT crystal substrate at 700~$^oC$ in an off-axis sputtering system. The deposition pressure was kept at 0.7~mbar by flowing a mix of Argon and Oxygen gas (3:2). The sputtering power was 50~W. Still at the deposition pressure, the system was cooled down to room temperature. At first, the cooling rate is 25~$^oC$/min, but we waited several hours to reach a low temperature and minimize possible issues with interdiffusion. The epitaxy of the LSMO film was confirmed before \cite{jungxiang2023triplet}. Then the system was pumped down to 10$^{-7}$~mbar, followed by the sputter deposition of NbTi (60~nm) on the LSMO layer $in~situ$, in an Ar atmosphere with a pressure of 0.1~mbar.

Next, a positive e-beam resist was spin-coated on the NbTi/LSMO bilayer, and a pattern was written by exposure to an electron beam (Fig.~\ref{lrp-fig1}a-I, II). The etch rate of resist is high in ion beam etching (IBE), and the resist mask will be removed completely before succeeding to pattern the NbTi/LSMO bilayer. Therefore, the sample with the patterned resist layers was loaded into an on-axis radio-frequency sputtering system, and a thick Pt layer (426 nm) was deposited. After lifting off, the Pt served as a hard mask protecting the NbTi/LSMO bilayer (Fig.~\ref{lrp-fig1}a-III). In the next step, shown in Fig.~\ref{lrp-fig1}a-IV, the NbTi/LSMO bilayer was patterned. In this step, the Pt layer is (almost) fully removed, but the bilayer should be intact. To open up a gap and form a junction, \ie making LSMO the weak link, the NbTi needs to be removed. For this, we repeated the above procedure with Pt liftoff. The difference in this step was that the central part of the nanowire was uncovered by the Pt hard mask and hence the IBE removed the NbTi in that part (Fig.~\ref{lrp-fig1}a-V). Crucial here is that the thickness of the Pt layer is chosen such that it is removed during the same etch time that removes the 60~nm NbTi layer. In this way, we obtained well-defined lateral NbTi/LSMO/NbTi junctions with adjustable widths and lengths (Fig.~\ref{lrp-fig1}a-VI).

As mentioned, several etch steps in the procedure depend crucially on the etch rate for the different materials. These were collected in a series of precise measurements on films of the relevant materials, involving thickness measurements by small-angle x-ray diffraction and Atomic Force Microscopy. Here we summarize details of the etch process and give the etching rates that we found and used in step IV (removing the Pt mask from the full structure), and step VI (removing the NbTi on top of the LSMO bridge). Thicknesses, etch rates, and etch times are given in Table~\ref{lrp-etching}.
Subsequent to the nanofabrication, we employed energy-dispersive X-ray spectroscopy (EDS) to qualitatively examine the NbTi/LSMO junctions. We mapped the element Nb, as shown in Fig.~\ref{lrp-fig1}b. Before etching, a relatively strong EDS signal was observed Nb (left image in Fig.~\ref{lrp-fig1}b), which was absent after optimal etching (right image in Fig.~\ref{lrp-fig1}b). Given the resolution of a few nm, this does not prove we fully removed the Nb, but it shows that we did etch Nb in the intended area. In all, we fabricated four measurable devices in this way, two with bridge widths around 2~$\mu$m (called DV2 and DV3), and three with bridge widths of about 5~$\mu$m (called DV4, DV5, DV6). Device DV2 had a too low T$_c$ to be measured, DV5 was used for etch tests, and DV6 could only be measured at one temperature because of a cryostat problem. Both DV3 and DV4 showed the desired LRP effects. Below, we discuss the data on DV4 and DV5 as the purpose of the paper is to establish proof-of-principle regarding the device fabrication. \\

\begin{figure*}[ht]
\centering
\includegraphics[width=.7\textwidth]{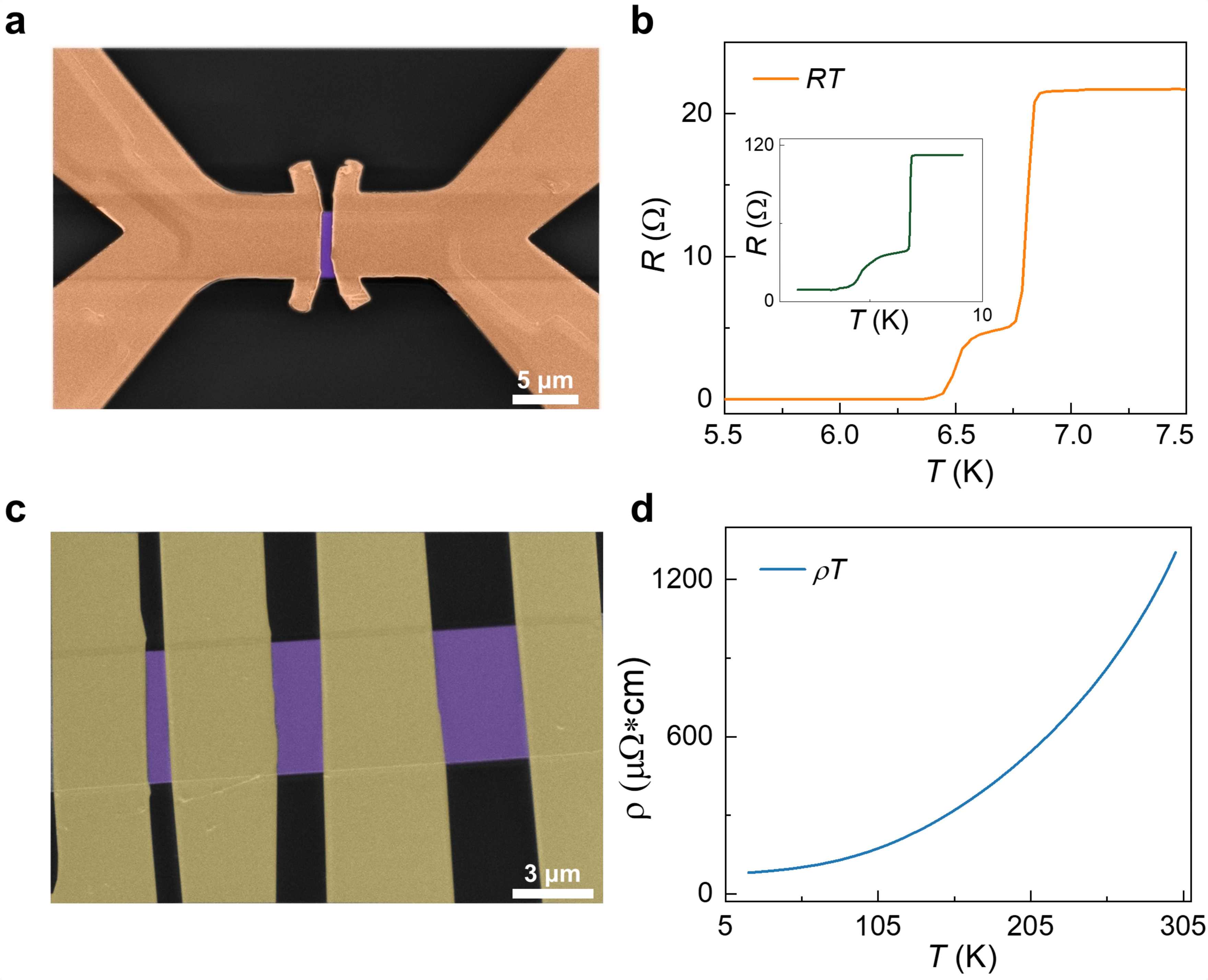}
\caption{\label{lrp-fig2} Electrical measurements. (a) Top-view SEM image of the junction with a length of 0.76~$\mu$m. (b) $R$-$T$ measurements after the optimal etching and over etching (inset), respectively. (c) SEM image of an LSMO nanowire with Pt leads in a four-probe geometry. (d) $\rho-T$ measurement showing the resistivity of the LSMO nanowire with a width of 5~$\mu$m is 80~$\mu \Omega$cm at 10~K.}
\end{figure*}
%
\vspace{3mm}

The electrical transport measurements also lead to the conclusion that the etching has removed the NbTi layer, and that an LSMO-based junction is formed. First we measured the temperature-dependent resistance ($R$-$T$) of junction DV5, using a current of 1~$\mu$A (as in all $R$-$T$ measurements). The length of the junction and the width of the central bridge are determined to be 0.766~$\mu$m and 4.89~$\mu$m, respectively, by scanning electron microscopy (SEM) (Fig.~\ref{lrp-fig2}a). The $R$-$T$ data after the etching are shown in Fig.~\ref{lrp-fig2}b (orange curve). We observe the well-known signature of a proximized structure, where the banks become superconducting first, followed by a plateau that is determined by the resistance of the weak link, and then a resistance drop to zero when the link becomes fully superconducting. The first drop is seen around 6.9~K, the second one below 6.5~K. The relatively low T$_c$ of the NbTi is due to a background pressure of 10$^{-7}$~mbar and the fact that NbTi is a good getter. \\

\noindent
We can approximately calculate the (expected) resistance of the junction in case the weak link consists of LSMO. The resistivity $\rho$ of a typical LSMO nanowire with a width of 5~$\mu$m was measured using the four-probe method, as shown in Fig.~\ref{lrp-fig2}c. The $\rho$ is determined to be 80~$\mu \Omega$cm at 10~K (Fig.~\ref{lrp-fig2}d). According to $R_{cal} = \rho l/S$, $l$ is the length, and $S$ is the cross-section area, we obtain $R_{cal} \approx 3.1~\Omega$. The measured resistance of the junction is about 5~$\Omega$, slightly larger than the calculated value, indicating the LSMO layer is a bit over-etched. Further etching leads to an 8 times increase in the resistance (inset in Fig.~\ref{lrp-fig2}b) since now also the NbTi of the contacts is being etched.  Since IBE is utilized to remove the unwanted NbTi layer, it may also contaminate the LSMO layer if the etching time is not optimized, leading to a pronounced increase in resistance. Therefore, we take the optimal etching to make the NbTi/LSMO junctions to ensure the NbTi is completely removed, and LSMO is slightly etched to form a ferromagnetic weak link. \\

\begin{figure*}[ht]
\centering
\includegraphics[width=.8\textwidth]{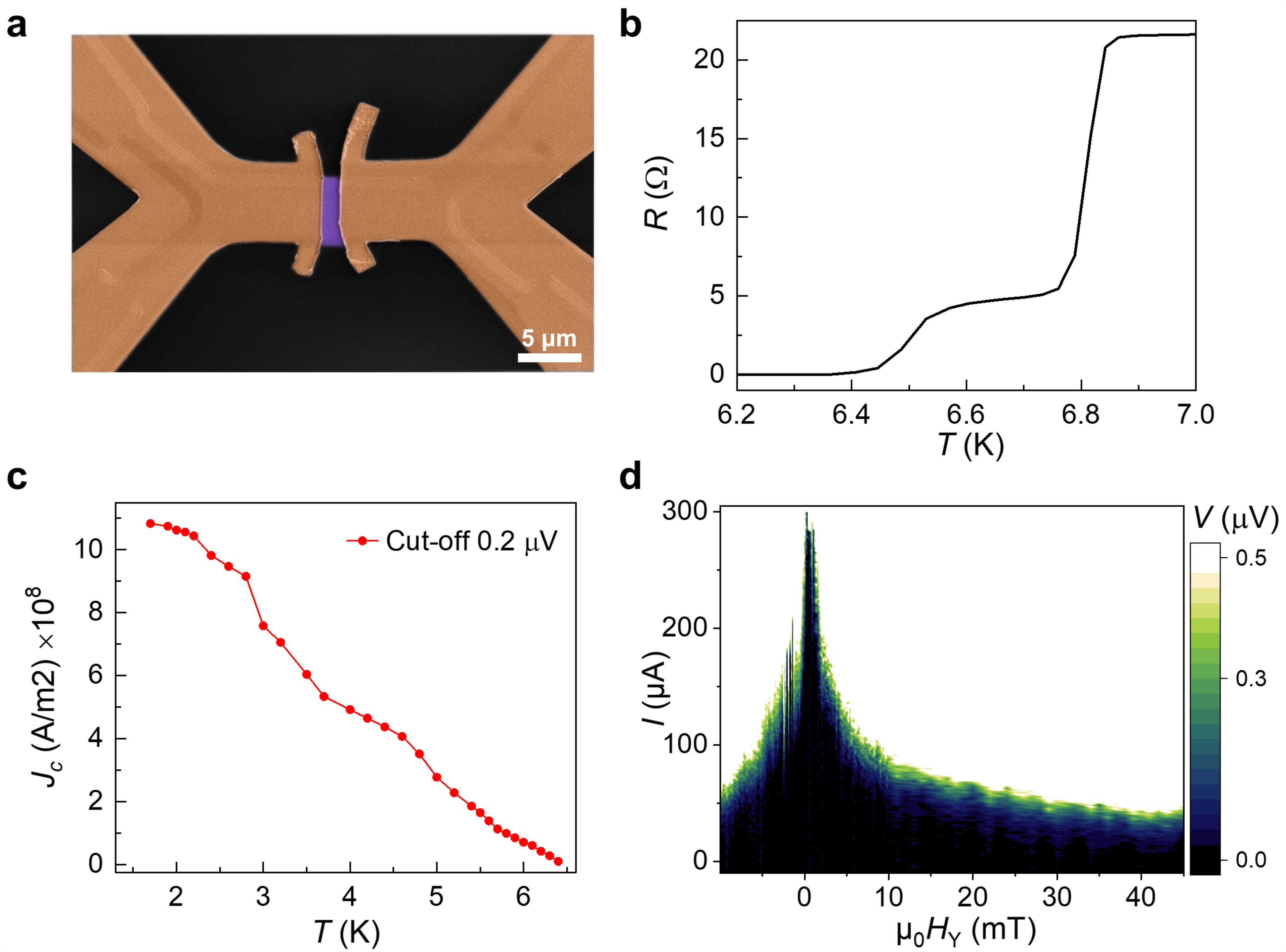}
\caption{\label{lrp-fig3} Transport characterizations of the junction with a length of 1.3~$\mu$m. (a) False-colored SEM image. (b) A plot of the $R-T$ measurement after the optimal etching. (c) A plot of the temperature-dependent $J_c$ extracted by taking a criterion of 0.2~$\mu$V. (d) SQI pattern obtained by sweeping $B_\perp$ ($y$ direction) at 6.3~K.}
\end{figure*}

\noindent

\begin{figure*}[ht]
\centering
\includegraphics[width=.8\textwidth]{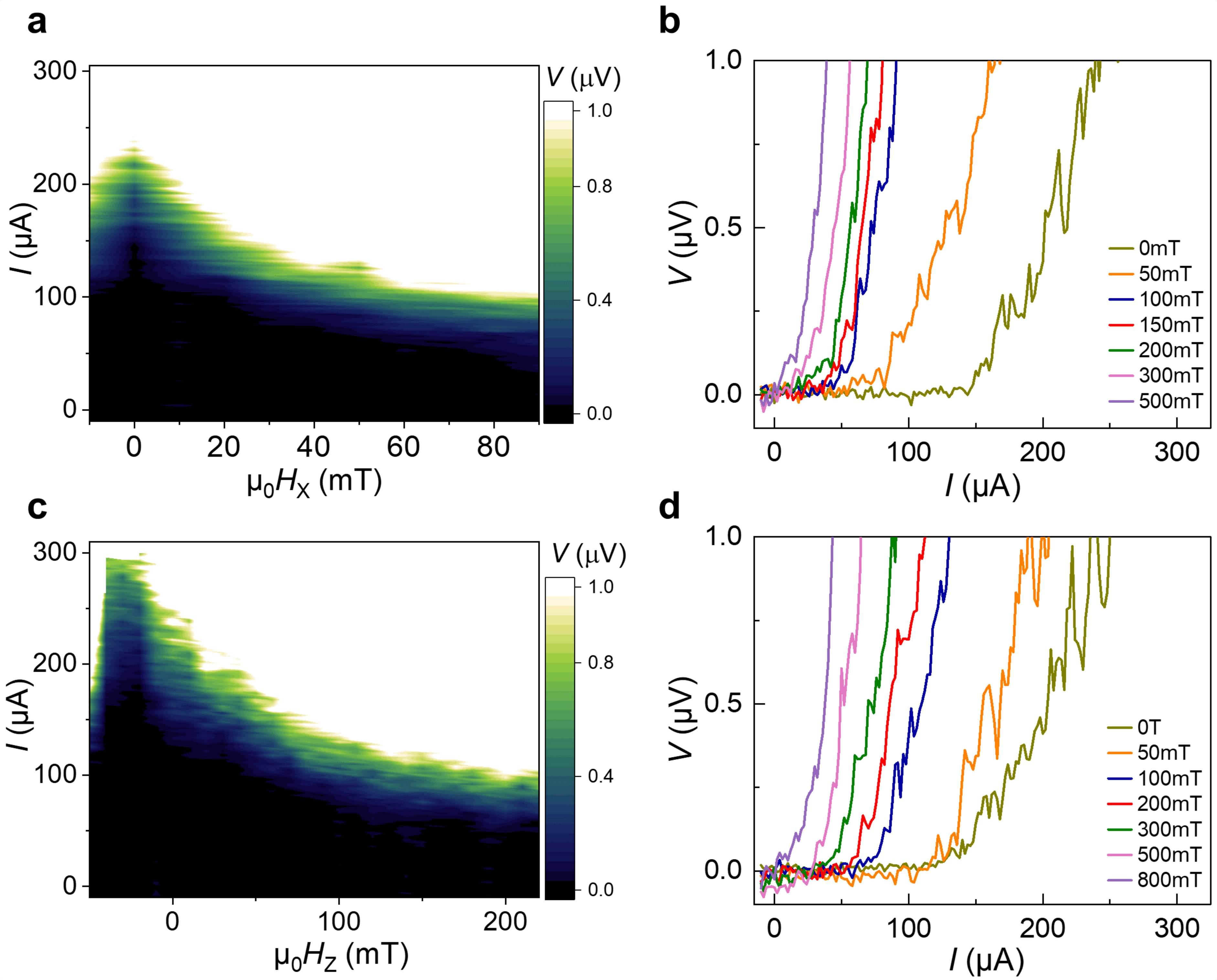}
\caption{\label{lrp-fig4} Transport measurements with $B_\parallel$. (a) SQI pattern obtained by sweeping $B_\parallel$ along $x$ direction, \ie the long axis of the wire. (b) Extracted $IV$ curves at different fields along $x$. (c) SQI pattern measured by sweeping $B_\parallel$ along $z$ direction, \ie the short axis of the wire, and (d) corresponding extracted $IV$ curves.}
\end{figure*}

Having fabricated the lateral NbTi/LSMO junctions, we performed transport measurements on junction DV4 with 1.3~$\mu$m length ($l$) and 5~$\mu$m width ($w$) (Fig.~\ref{lrp-fig3}a), by changing temperature and sweeping fields, yielding $R-T$ and SQI characterizations. The $R-T$ curve was plotted in Fig.~\ref{lrp-fig3}b. The resistance of the junction decreases to zero after two transitions. The first transition corresponds to the superconducting NbTi, leaving a residual resistance of 6~$\Omega$. We take $\rho \approx$~80~$\mu \Omega$cm, $R_{cal} \approx$~5.2~$\Omega$, meaning LSMO is over-etched optimally in this junction. The second transition signifies the onset of the superconductivity in the weak link, indicating LSMO is fully proximized. The LRP effect over such a long length is not inconceivable. Sanchez-Manzano \etal reported the LRP effect in very wide YBCO/LSMO junctions (20 and 25~$\mu$m), with a length of about 1~$\mu$m, although with much lower critical currents \cite{sanchez2022extremely}. Therefore, we thoroughly inspected the observed superconductivity in the junction here. The behavior of the critical current density ($J_c$) versus temperature was measured first and summarised in Fig.~\ref{lrp-fig3}c by taking a criterion of 0.2~$\mu$V. At the base temperature (1.5~K), $J_c$ is about 11$\times$10$^8$~A/m$^2$, which is a really high value over such as long length.

To further confirm the Josephson coupling, we examined the SQI pattern of the junction by applying a magnetic field $B_{\perp}$ ($y$ direction; perpendicular to the sample plane) at 6.3~K (Fig.~\ref{lrp-fig3}d). The obtained SQI pattern is far from the typical Fraunhofer-like pattern. Instead, it is quasi-Gaussian-like with a rather sharp decay of $I_c (B)$. If we ignore the shielding currents in the side NbTi arms, the first minimum for the Fraunhofer pattern would be expected at $\Delta B = \Phi_0/(wl)$. This yields $\Delta B \approx$~0.3~mT. Therefore, the expected period is very small (actually smaller than can be easily measured), but the observed pattern is wider and falls off more slowly. Note that, from the measured resistivity of LSMO (40~$\mu\Omega$cm), its mean free path ($\ell_H$) is about 6.5~nm \cite{nadgorny2001}. With a Fermi velocity of
$v_F \sim 7.4\times10^5$~m/s and using $T_c$~= 6.4~K, we find the diffusive coherence length $\xi_{HMF} = \sqrt{\hbar (v_F \ell_H/3) /(2\pi k_B T_c)}$ to be $\approx$~17.5~nm. Given the length of the junction of about 1.3~$\mu$m, the case we deal with, for the triplet transport, is that of a diffusive long junction  ($\ell_H \ll \xi_{HMF} \ll l$). That can lead to a Gaussian-shaped curve of the form $I_c (B)/I_c(0) = e^{-\Phi^2/(2\sigma^2\Phi_0^2)}$ under the condition that $L/w \gg 1$ \cite{cuevas2007,chiodi2012geometry, barone1982physics}. Here $\Phi$ is the total flux threading the junction, $\sigma$ is a geometry-dependent constant, and $\Phi_0$ is the flux quantum. The problem is that for this junction, $l/w \approx 0.26$, which should lead to a Fraunhofer pattern. Here we have to remember that we are dealing with a magnetic structure with internal magnetic fields. The phase accumulation along paths, \ie along the width or along the length of the wire may well be different. A more quantitative description has not been made yet \cite{suominen2017anomalous,borcsok2019fraunhofer,alidoust2018fraunhofer}.

Afterward, we recorded the SQI patterns with $B_\parallel$ (field in plane along the wire) at 6.3~K, as shown in Fig.~\ref{lrp-fig4}. With $B_\parallel$ along the long axis ($x$ direction) or short axis ($z$ direction) of the wire, the obtained SQI patterns behave quite differently. The central peaks are much wider than that of the SQI pattern with $B_\perp$. Fig.~\ref{lrp-fig4}a,c show the critical current decays far more gradually. The $I-V$ curves measured at different fields demonstrate $I_c$ vanishes at the fields of the order of hundreds of mT (Fig.~\ref{lrp-fig4}b and d). The $I-V$ curves become nearly linear at 500~mT (along $x$) and 800~mT (along $z$). Note that the $x$ direction is the current direction, and in the configuration B $\parallel x$, no SQI is to be expected. \\


To summarize, we developed a nanofabrication method to fabricate long lateral NbTi/LSMO Josephson junctions. The process starts with bilayer films and involves using a hard mask of Pt, and a series of lithography steps. One crucial step is the removal of the 60~nm NbTi layer on the LSMO bridge. By slightly overetching, the EDS and resistance measurements confirm that LSMO-based junctions have been made, with absence of NbTi shorts. We measured and characterized a junction with 1.3~$\mu$m length and 5~$\mu$m width by measuring the temperature dependence of $J_c$ and recording the SQI patterns with either $B_\perp$ or $B_\parallel$. The junction has a critical current of 11~$\times$10$^8$~A/m$^2$ at 1.5~K. Distinguishing Gaussian-shaped SQI patterns are observed with the fields along different directions. We fabricated several devices that showed a full proximity effect in their $R-T$ characteristics. What still needs to be improved is understanding of, and control over, the triplet generator. As to control, a device (DV3) with a smaller bridge width (2~$\mu$m), and a smaller length (0.59~$\mu$m) showed a smaller value of $J_c$ at 1.5~K (5~$\times$10$^8$~A/m$^2$) than the longer device mentioned above. This length dependence is the focus of current experiments, the more so since the results for DV4 hint at the possibility of still longer lengths. With respect to understanding, similar to the YBCO/LSMO case, we did not build in an extra magnetic layer to enhance spin scattering. Preliminary transmission electron microscopy data indicate that the oxygen concentration at our NbTi/LSMO interface is anomalous, which might give rise to inhomogeneous magnetism \cite{mariona}. We surmise that these long junctions will open the way to further producing spin torque effects and manipulating local spins or even moving domain walls without Joule heating, giving rise to actual applications in superconducting spintronics.

\section*{Acknowledgments}
The authors thank Kaveh Lahabi and Remko Fermin for discussions, and Marcel Hesselberth and Douwe Scholma for support in device preparation. Y.J. was partially funded
by China Scholarship Council (Grant No. 201808440424). The work is partly financed by the Dutch Research Council (NWO) through Projectruimte Grant No. 680.91.128 and Project OCENW.XS22.2.032. The work was further supported by EU Cost Action CA21144 (SUPERQUMAP).

\section*{Author Declarations}
\subsection*{Conflicts of interest}
The authors declare no competing interests

\subsection*{Author contributions}
\textbf{Yunxiang Yao}: Conceptualization (equal); Data curation (lead); Formal analysis (lead); Funding acquisition (supporting); Investigation (lead); Methodology (equal); Project administration (support); Resources (lead); Software (lead); Supervision (supporting); Validation (equal); Visualization (lead); Writing – original draft (lead); Writing – review \& editing (equal). \textbf{Jan Aarts}: Conceptualization (equal); Data curation (supporting); Formal analysis (supporting); Funding acquisition (lead); Investigation (supporting); Methodology (equal); Project administration (lead); Resources (supporting); Software (supporting); Supervision (lead); Validation (equal); Visualization (supporting); Writing – original draft (supporting); Writing – review \& editing (equal).

\subsection*{Data availability}
The data that support the findings of this study are available from the corresponding authors upon reasonable request.


\bibliographystyle{unsrtnat}

\begin{thebibliography}{16}
\providecommand{\natexlab}[1]{#1}
\providecommand{\url}[1]{\texttt{#1}}
\expandafter\ifx\csname urlstyle\endcsname\relax
  \providecommand{\doi}[1]{doi: #1}\else
  \providecommand{\doi}{doi: \begingroup \urlstyle{rm}\Url}\fi

\bibitem[Keizer et~al.(2006)Keizer, G{\"o}nnenwein, Klapwijk, Miao, Xiao, and Gupta]{keizer2006spin}
Ruurd~S Keizer, Sebastian~TB G{\"o}nnenwein, Teun~M Klapwijk, Guoxing Miao, Gang Xiao, and Arunava Gupta.
\newblock A spin triplet supercurrent through the half-metallic ferromagnet {C}r{O}$_2$.
\newblock \emph{Nature}, 439\penalty0 (7078):\penalty0 825--827, 2006.

\bibitem[Anwar et~al.(2010)Anwar, Czeschka, Hesselberth, Porcu, and Aarts]{anwar2010long}
MS~Anwar, F~Czeschka, M~Hesselberth, M~Porcu, and J~Aarts.
\newblock Long-range supercurrents through half-metallic ferromagnetic {C}r{O}$_2$.
\newblock \emph{Physical Review B}, 82\penalty0 (10):\penalty0 100501, 2010.

\bibitem[Singh et~al.(2016)Singh, Jansen, Lahabi, and Aarts]{singh2016high}
Amrita Singh, Charlotte Jansen, Kaveh Lahabi, and Jan Aarts.
\newblock High-quality cro 2 nanowires for dissipation-less spintronics.
\newblock \emph{Physical Review X}, 6\penalty0 (4):\penalty0 041012, 2016.

\bibitem[Sanchez-Manzano et~al.(2022)Sanchez-Manzano, Mesoraca, Cuellar, Cabero, Rouco, Orfila, Palermo, Balan, Marcano, Sander, et~al.]{sanchez2022extremely}
D~Sanchez-Manzano, S~Mesoraca, FA~Cuellar, M~Cabero, V~Rouco, G~Orfila, X~Palermo, A~Balan, L~Marcano, A~Sander, et~al.
\newblock Extremely long-range, high-temperature josephson coupling across a half-metallic ferromagnet.
\newblock \emph{Nature Materials}, 21\penalty0 (2):\penalty0 188--194, 2022.

\bibitem[Anwar and Aarts(2011)]{2011anwar}
MS~Anwar and J~Aarts.
\newblock Inducing supercurrents in thin films of ferromagnetic cro2.
\newblock \emph{Superconductor Science and Technology}, 24\penalty0 (2):\penalty0 024016, 2011.

\bibitem[Visani et~al.(2012)Visani, Sefrioui, Tornos, Leon, Briatico, Bibes, Barth{\'e}l{\'e}my, Santamaria, and Villegas]{visani2012equal}
C~Visani, Z~Sefrioui, J~Tornos, C~Leon, J~Briatico, M~Bibes, A~Barth{\'e}l{\'e}my, J~Santamaria, and Javier~E Villegas.
\newblock Equal-spin andreev reflection and long-range coherent transport in high-temperature superconductor/half-metallic ferromagnet junctions.
\newblock \emph{Nature Physics}, 8\penalty0 (7):\penalty0 539--543, 2012.

\bibitem[Chakhalian et~al.(2006)Chakhalian, Freeland, Srajer, Strempfer, Khaliullin, Cezar, Charlton, Dalgliesh, Bernhard, Cristiani, et~al.]{chakhalian2006magnetism}
Jak Chakhalian, JW~Freeland, G~Srajer, J~Strempfer, G~Khaliullin, JC~Cezar, T~Charlton, R~Dalgliesh, Ch~Bernhard, G~Cristiani, et~al.
\newblock Magnetism at the interface between ferromagnetic and superconducting oxides.
\newblock \emph{Nature Physics}, 2\penalty0 (4):\penalty0 244--248, 2006.

\bibitem[Jungxiang et~al.(2023)Jungxiang, Fermin, Lahabi, and Aarts]{jungxiang2023triplet}
Yao Jungxiang, Remko Fermin, Kaveh Lahabi, and Jan Aarts.
\newblock Triplet supercurrents in lateral josephson junctions with a half-metallic ferromagnet, 2023.
\newblock URL \url{https://arxiv.org/abs/2303.13922}.

\bibitem[Nadgorny et~al.(2001)Nadgorny, Mazin, Osofsky, Soulen, Broussard, Stroud, Singh, Harris, Arsenov, and Mukovskii]{nadgorny2001}
B.~Nadgorny, II. Mazin, M.~Osofsky, RJ~Soulen, P.~Broussard, RM~Stroud, DJ. Singh, V.G. Harris, A.~Arsenov, and Y.~Mukovskii.
\newblock Origin of high transport spin polarization in {L}a$_{0.7}${S}r$_{0.3}${M}n{O}$_3$: Direct evidence for minority spin states.
\newblock \emph{Physical Review B}, 63:\penalty0 184433, 2001.

\bibitem[Cuevas and Bergeret(2007)]{cuevas2007}
J.~C. Cuevas and F.~S. Bergeret.
\newblock Magnetic interference patterns and vortices in diffusive sns junctions.
\newblock \emph{Physical Review Letters}, 99:\penalty0 217002, 2007.

\bibitem[Chiodi et~al.(2012)Chiodi, Ferrier, Gu{\'e}ron, Cuevas, Montambaux, Fortuna, Kasumov, and Bouchiat]{chiodi2012geometry}
F~Chiodi, M~Ferrier, S~Gu{\'e}ron, JC~Cuevas, G~Montambaux, F~Fortuna, A~Kasumov, and H~Bouchiat.
\newblock Geometry-related magnetic interference patterns in long s n s josephson junctions.
\newblock \emph{Physical Review B}, 86\penalty0 (6):\penalty0 064510, 2012.

\bibitem[Barone and Paterno(1982)]{barone1982physics}
Antonio Barone and Gianfranco Paterno.
\newblock \emph{Physics and applications of the Josephson effect}, volume~1.
\newblock Wiley Online Library, 1982.

\bibitem[Suominen et~al.(2017)Suominen, Danon, Kjaergaard, Flensberg, Shabani, Palmstr{\o}m, Nichele, and Marcus]{suominen2017anomalous}
HJ~Suominen, J~Danon, M~Kjaergaard, K~Flensberg, J~Shabani, CJ~Palmstr{\o}m, F~Nichele, and CM~Marcus.
\newblock Anomalous fraunhofer interference in epitaxial superconductor-semiconductor josephson junctions.
\newblock \emph{Physical Review B}, 95\penalty0 (3):\penalty0 035307, 2017.

\bibitem[B{\"o}rcs{\"o}k et~al.(2019)B{\"o}rcs{\"o}k, Komori, Buzdin, and Robinson]{borcsok2019fraunhofer}
Bence B{\"o}rcs{\"o}k, Sachio Komori, AI~Buzdin, and JWA Robinson.
\newblock Fraunhofer patterns in magnetic josephson junctions with non-uniform magnetic susceptibility.
\newblock \emph{Scientific Reports}, 9\penalty0 (1):\penalty0 1--6, 2019.

\bibitem[Alidoust et~al.(2018)Alidoust, Willatzen, and Jauho]{alidoust2018fraunhofer}
Mohammad Alidoust, Morten Willatzen, and Antti-Pekka Jauho.
\newblock Fraunhofer response and supercurrent spin switching in black phosphorus with strain and disorder.
\newblock \emph{Physical Review B}, 98\penalty0 (18):\penalty0 184505, 2018.

\bibitem[mar()]{mariona}
M. Cabero Piris, private communication.

\end{thebibliography}

\end{multicols}
\end{document}